\documentclass[twocolumn,showpacs,preprintnumbers,amsmath,amssymb]{revtex4}

\usepackage{graphicx}
\usepackage{dcolumn}
\usepackage{bm}
\newcommand{\vv}[1]{\mbox{\boldmath{$#1$}}}

\begin{document}

\title{Electron Transport in Magnetic-Field-Induced Quasi-One-Dimensional\\Electron Systems in Semiconductor Nanowhiskers}

\author{Toshihiro Kubo}
 \affiliation{Department of Physics, Tokyo University of Science.}
 \altaffiliation[Also at ]{ICORP, JST.\\}
 
 \email{kubo@tarucha.jst.go.jp}
 
 \author{Yasuhiro Tokura}
 
\affiliation{%
NTT Basic Research Laboratories, NTT Corporation
}%
\email{tokura@will.brl.ntt.co.jp}
\date{\today}

\begin{abstract}
Many-body effects on tunneling of electrons in semiconductor nanowhiskers are investigated in a magnetic quantum limit. We consider the system with which bulk and edge states coexist. We show that interaction parameters of edge states are much smaller than those of bulk states and the tunneling conductance of edge states hardly depends on temperature and the singular behavior of tunneling conductance of bulk states can be observed.
\end{abstract}

\maketitle

\section{Introduction}

The electrons in an isotropic bulk conductor under a very strong magnetic field provides an interesting example of a quasi-one-dimensional electron system. It is possible to imagine a case where all electrons are accommodated in the lowest Landau subband in such a \textit{magnetic quantum limit} (MQL). We thus expect its transport properties to be similar to those of one-dimensional (1D) electron systems. Many-body effects on the electron transport in the magnetic-field-induced quasi-1D electron systems have recently been investigated \cite{maslov,glazman,kubo}. In such systems, measurement of the electron transport is much easier than in 1D electron systems, since it can be performed with the use of a bulk specimen. The tunneling conductance is qualitatively similar to that of a 1D Tomonaga-Luttinger liquid (TLL) \cite{matveev,kawabata}, except that the parameter of electron-electron interaction is magnetic field dependent, and may be either positive or negative.

Those theoretical predictions are experimentally verifiable by low-carrier-density materials such as doped semiconductors. In order to observe interaction effects clearly, in a MQL, the mean free path of electrons has to be much longer than the Fermi wavelength. However, for bulk-doped semiconductors, it is difficult to satisfy that condition. Therefore, we consider semiconductor nanowhiskers as more realistic systems since an  extremely high carrier mobility is expected in modulation doped structures. Recently, high-quality semiconductor nanowhiskers with sharp heterojunctions have been realized \cite{whisker1,whisker2}. We study many-body effects on the tunneling conductance in nanowhiskers whose radii are much longer than the Larmor radius. In such systems, both bulk and edge states may contribute to electron transport. In Ref. \cite{kubo2}, we showed that the contributions of edge states to the tunneling conductance of bulk states can be neglected within the approximation which takes into account only most divergent terms at low temperatures. However, it was also shown that there can be a situation where the tunneling of electrons in edge states may not be ignored experimentally.

Therefore, in this report, we study numerically the effects of electron-electron interaction on the tunneling of electrons in edge states. It will be shown that interaction parameters of edge states are much smaller than those of bulk states.

\section{Model}
We consider the semiconductor nanowhisker whose radius is much longer than the Larmor radius $\lambda_B=\sqrt{\hbar/eB}$. We ignore the spin degree of freedom, assuming that spin of every electron is completely polarized in the direction of the magnetic field $B$. We choose the z-axis of the coordinate frame along the magnetic field applied parallel to the growth direction of the nanowhiskers, and the symmetric gauge $\vec{A}=(-By/2,Bx/2,0)$ for the vector potential $\vec{A}$ is used.
\begin{figure}[h]
\begin{center}\leavevmode
\includegraphics[width=0.5\linewidth]{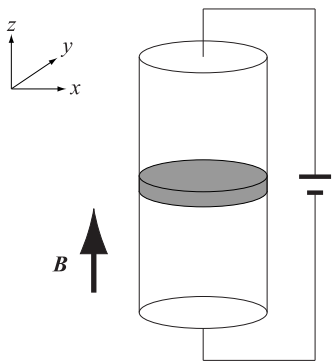}
\caption{Tunnel junction in semiconductor nanowhisker with radius much longer than Larmor radius. The magnetic field $\vv{B}=(0,0,B)$ is applied normal to the tunnel barrier.}\label{nanowhisker}\end{center}\end{figure}

A tunnel junction in a nanowhisker (see Fig. \ref{nanowhisker}) is described by the following Hamiltonian
\begin{equation}
H_0=\frac{\left(\vv{p}+e\vv{A}(\vv{x}) \right)^2}{2m}+W(x,y)+U(z).\label{hamiltonian}
\end{equation}
Here, $W(x,y)$ is the confining potential of a nanowhisker which is approximated by the infinite square well
\begin{align}
W(x,y)=\left\{
  \begin{array}{cc}
    0,   &  \sqrt{x^2+y^2}<R  \\
    \infty,   &  \sqrt{x^2+y^2}>R  \\
  \end{array}
\right.
,\label{potential}
\end{align}
where $R$ is the radius of the nanowhisker, $U(z)$ is the barrier potential, and the potential barrier is assumed to be localized around $z=0$, i.e., $U(z)=0$ for $|z|>a$.

We consider the case where a MQL is realized. In such a case the magnetic field is sufficiently strong that
\begin{equation}
B>\frac{\hbar}{e}(2\pi^4{n_e}^2)^{1/3},\label{mql}
\end{equation}
where $n_e$ is the electron density. Moreover, the wave functions and energy eigenvalues of edge states are defined by $\chi_{\ell,k_z}(\vec{x})=f_{\ell}(r)e^{i\ell\theta}u_{k_z}(z)$ and $\epsilon_{\ell,k_z}$. Here $\ell$ is the angular momentum quantum number and the lowest Landau subband corresponds to $\ell\le 0$. Then, we have effective Schr\"{o}dinger equation of radial direction with the angular momentum quantum number $\ell$,
\begin{eqnarray}
& &\left[-\frac{\hbar^2}{2m}\left(\frac{\partial^2}{\partial r^2}+\frac{1}{r}\frac{\partial}{\partial r}-\frac{{\ell}^2}{r^2} \right)\right.\nonumber\\
& &\quad \left.+\frac{1}{2}\hbar\omega_c\ell+\frac{{m}^2{\omega_c}^2}{4\hbar^2}r^2 \right]f_{\ell}(r)=\epsilon_{\ell,\perp}f_{\ell}(r),
\end{eqnarray}
where $\omega_c=eB/m$ is the cyclotron frequency and $\epsilon_{\ell,\perp}=\epsilon_{\ell,k_z}-\hbar^2{k_z}^2/2m$. This equation should be solved under the two boundary conditions: $\left.\nabla f_{\ell}(r)\right|_{r=0}=0$ and $\left.f_{\ell}(r)\right|_{r=R}=0$. Unfortunately, there is no analytical solution and we treat the edge states numerically.

\begin{figure}[h]
\begin{center}\leavevmode
\includegraphics[width=0.7\linewidth]{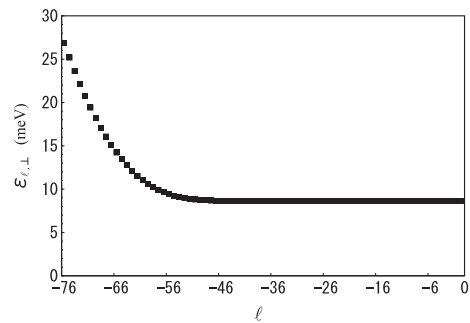}
\caption{Energy spectrum of the lowest Landau subband in the case of $R=100$ nm, $n_e=10^{23}$ $m^{-3}$, and $B=10$ tesla for GaAs.}\label{edge-energy}\end{center}\end{figure}

As an example, in the case of $R=100$ nm, $n_e=10^{23}$ $m^{-3}$, and $B=10$ tesla for GaAs, the energy spectrum of the lowest Landau subband is shown in Fig. \ref{edge-energy}. In this case, Fermi energy $\epsilon_F$ is $18.26$ meV $(\ell\simeq -70)$ and the position of a hard wall of the confining potential corresponds to $\ell\simeq -76$.

\section{tunneling conductance of edge states}
Since the Fermi wave number of the edge states differs between all modes, only when the electron in the edge state is scattered on the density modulation due to its own mode does logarithmic singularity arise. Therefore, in this section, we investigate the tunneling conductance in a single edge mode (with only $\ell$).

\subsection{interaction parameter of edge states}
We calculate the tunneling conductance of an edge state with $\ell$ within the Hartree-Fock approximation using the same theoretical formulation in Ref. \cite{kubo}. Then, the tunneling conductance is written as
\begin{eqnarray}
G_{e,\ell}(T)=\frac{e^2}{h}\frac{\mathcal{T}_{0,\ell}\left(k_BT/E_{0,\ell} \right)^{2\alpha_{e,\ell}(B)}}{\mathcal{R}_{0,\ell}+\mathcal{T}_{0,\ell}\left(k_BT/E_{0,\ell} \right)^{2\alpha_{e,\ell}(B)}}.\label{edge-tunneling}
\end{eqnarray}
Note that Eq. (\ref{edge-tunneling}) is valid at $T<T_{F,\ell}$ \cite{kubo}. Here $T_{F,\ell}\equiv\hbar^2{k_{F,\ell}}^2/2m$ is the effective Fermi temperature, $\mathcal{T}_{0,\ell}$ is the bare transmission probability of an edge state with $\ell$, $\mathcal{R}_{0,\ell}=1-\mathcal{T}_{0,\ell}$, $E_{0,\ell}$ is the effective bandwidth, and
\begin{eqnarray}
\alpha_{e,\ell}(B)&=&\frac{1}{8\pi^3\hbar v_{F,\ell}}\int r_1dr_1\int_{0}^{2\pi}d\theta_1\int r_2dr_2\int_{0}^{2\pi}d\theta_2\nonumber\\
& &\times\left|f_{\ell}(r_1) \right|^2\left|f_{\ell}(r_2) \right|^2\nonumber\\
& &\times\left[\tilde{V}_{s,e}(r_{12};0)-\tilde{V}(r_{12};2k_{F,\ell}) \right],
\end{eqnarray}
where $r_{12}=\sqrt{{r_1}^2+{r_2}^2-2r_1r_2\cos(\theta_1-\theta_2)}$, $k_{F,\ell}$ is the Fermi wave number of an edge state with $\ell$, and $\tilde{V}(r_{12};k_z)$ is the Fourier transform of the bare Coulomb interaction potential with respect only to $z$ component defined by
\begin{eqnarray}
\tilde{V}(r_{12};k_z)&\equiv&\int^{\infty}_{-\infty}dzV(r_{12},z)e^{-ik_zz}\nonumber\\
&=&\frac{e^2}{2\pi\epsilon}K_0\left(k_zr_{12} \right),
\end{eqnarray}
where $K_0(r)$ is the modified Bessel function. $\tilde{V}_{s,e}(r_{12};k_z)$ is the Fourier transform of the screened Coulomb interaction potential with respect only to $z$ component defined in \S \ref{screen}. 

\subsection{Screened Coulomb Interaction in Nanowhisker}\label{screen}
The linearized Poisson's equation with Thomas-Fermi approximations is given by
\begin{equation}
\nabla^2\Phi(\vec{x})=\kappa^2(\vec{x})\cdot\Phi(\vec{x}),\label{poisson}
\end{equation}
where in our nanowhisker model
\begin{equation}
\kappa^2(\vec{x})=\sum_{\ell}\frac{e^2}{2\pi^2\epsilon\hbar v_{F,\ell}}\left|f_{\ell}(r) \right|^2\equiv \kappa^2(r)
\end{equation}
is only the function of the distance from the center of whisker, $r$. $\kappa(r)$ is the inverse of the screening length.

From Eq. (\ref{poisson}), the Fourier transform $\Phi_q(\vec{r})$ of $\Phi(\vec{x})$ with respect only to $z$ component satisfies
\begin{equation}
\frac{1}{r}\frac{\partial}{\partial r}\left(r\frac{\partial\Phi_{q}(\vec{r})}{\partial r} \right)+\frac{1}{r^2}\frac{\partial^2\Phi_{q}(\vec{r})}{\partial\theta^2}=\left(q^2+\kappa^2(r) \right)\Phi_{q}(\vec{r}).
\end{equation}
Since we are interested in the behavior $q\to 0$ for the Fock term, we assume $q=0$.

In this report, for simplicity, we use local approximation, namely, the screening length is only the function of the electron densities at $r=r_1$ and $r=r_2$:
\begin{equation}
\tilde{V}_{s,e}(r_{12};0)=\frac{e^2}{2\pi\epsilon}K_0\left(\frac{\kappa(r_1)+\kappa(r_2)}{2}r_{12}\right).
\end{equation}

\subsection{Numerical Results}
\begin{figure}[h]
\begin{center}\leavevmode
\includegraphics[width=0.7\linewidth]{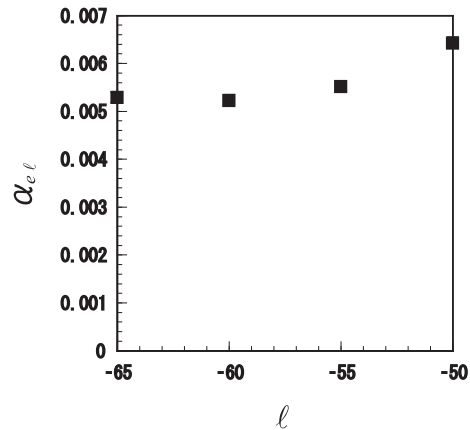}
\caption{Parameter of electron-electron interaction in edge states $\alpha_{e,\ell}(B)$ in the case of $R=100$ nm, $n_e=10^{23}$ $m^{-3}$, and $B=10$ tesla for GaAs.}\label{parameter}\end{center}\end{figure}

The parameters of electron-electron interaction in edge states in some case are shown in Fig. \ref{parameter}. We assume that radius of nanowhisker is $R=100$ nm and electron density is $n_e=10^{23}$ $m^{-3}$ in GaAs. Although there is no clear boundary between bulk and edge states, edge states with lower energy $\epsilon_{\ell,\perp}$ are not shown, and parameters in several representative edge states are shown in Fig. \ref{parameter}. As for states near the Fermi energy, the effective Fermi temperature $T_{F,\ell}$ is very low and the perturbative treatment for the electron-electron interaction is no longer valid because of the low electron density. Moreover, since Fermi wave length is very long and the mean free path is short because of the weak screening, we can ignore the scattering effects due to Freidel oscillation. Thus, states near the Fermi energy are also not shown in Fig. \ref{parameter}. In contrast, since the bulk states suffer scattering by the density modulation in all modes showing $2k_F$ oscillations, the relation of $|\alpha_{e,\ell}(B)|\ll |\alpha(B)|$ was predicted in Ref. \cite{kubo2}. In fact, in the case of $B=10$ tesla, the parameter $\alpha(B)\simeq 0.420$ in bulk states. From Fig. \ref{parameter}, in this case, we find $|\alpha_{e,\ell}(B)|\ll |\alpha(B)|$.

\begin{figure}[h]
\begin{center}\leavevmode
\includegraphics[width=0.9\linewidth]{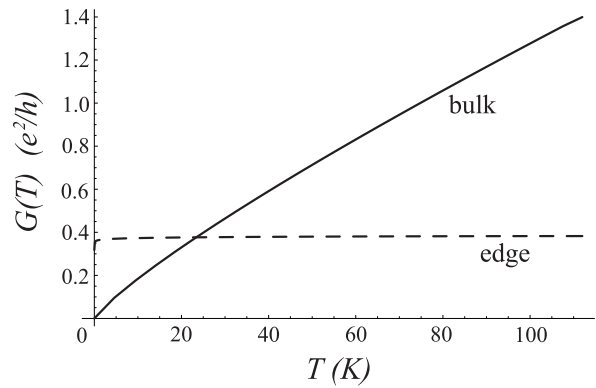}
\caption{Solid curve indicates the temperature dependence of the tunneling conductance of bulk states. Dashed curve indicates the temperature dependence of the tunneling conductance of edge states.}\label{conductance}\end{center}\end{figure}

Finally, we consider the temperature dependences of the tunneling conductance of bulk and edge states in the case of the realistic barrier potential $U(z)$ which has the form of the rectangular barrier defined by
\begin{eqnarray}
U(z)=U_0\theta\left(\frac{a}{2}-|z|\right)
\end{eqnarray}
with $U_0>0$. We assume that $a=2$ nm and $U_0=300$ meV under the realistic experimental conditions. Moreover, we assume that radius of nanowhisker is $R=100$ nm, electron density is $n_e=10^{23}$ $m^{-3}$ and magnetic field is $B=10$ tesla in GaAs. Under such conditions, the temperature dependences of the tunneling conductance of bulk and edge states are shown in Fig. \ref{conductance}. In this case, the effective Fermi temperature of bulk states is $T_F=112$ K.

Therefore, we expect that the singular behavior of the tunneling conductance of bulk states obtained in Refs. \cite{maslov,glazman,kubo} can be observed experimentally. Moreover, in Fig. \ref{conductance}, the magnitude of the tunneling conductance of edge states is larger than that of bulk states at $T<23.4$ K.

\section{Summary}
In summary, we have studied theoretically many-body effects on the tunneling of electrons through a barrier between the magnetic-field-induced quasi 1D electron systems. In particular, we investigated many-body effects on the tunneling of edge states numerically. The numerical results have been consistent with qualitative discussions in Ref. \cite{kubo2} and the absolute values of interaction parameters $|\alpha_{e,\ell}(B)|$ of edge states are much smaller than those of interaction parameters $|\alpha(B)|$ of bulk states.

\section*{Acknowledgements}
One of the authors (Y. T.) is partly supported by SORST-JST.

\end{document}